\documentclass[journal=jacsat,manuscript=article]{achemso}

\usepackage[version=3]{mhchem} 
\usepackage[T1]{fontenc}       
\usepackage{float}
\usepackage{graphicx}
\usepackage{hyperref}

\author{Denis Garoli}
\author{Pierfrancesco Zilio}

\author{Francesco De Angelis}
\email{francesco.deangelis@iit.it}

\affiliation[Istituto Italiano di Tecnologia]
{Istituto Italiano di Tecnologia, via Morego 30, I-16163, Genova, Italy}

\author{Yuri Gorodetski}
\email{yurig@ariel.ac.il}
\affiliation[Ariel University]
{Ariel University, Mechanical Engineering \& Mechatronics department and Electrical Engineering \& Electronics department, Ariel, 40700, Israel}
\title[Title]
  {Helicity locking in light emitted from a plasmonic nanotaper}

\keywords{Helicity, Plasmons, Nanophotonics, Spin-locking}

\begin{document}

%
%
%
%
%
%
\begin{abstract}
Surface plasmon waves carry an intrinsic transverse spin, which is locked to its propagation direction. Apparently, when a singular plasmonic mode is guided on a conic surface this spin-locking may lead to a strong circular polarization of the far--field emission. Specifically, an adiabatically tapered gold nanocone guides an a priori excited plasmonic vortex upwards where the mode accelerates and finally beams out from the tip apex. The helicity of this beam is shown to be single-handed and stems solely from the transverse spin-locking of the helical plasmonic wave-front. We present a simple geometric model that fully predicts the emerging light spin in our system. Finally we experimentally demonstrate the helicity-locking phenomenon by using accurately fabricated nanostructures and confirm the results with the model and numerical data.  
 
\end{abstract}

Surface plasmons (SP) propagating on nanoscale tapered conical waveguides were recently proposed as an optical emulation of a ``black hole'' as it features center-symmetric effective index increasing upon propagation\cite{BozhevolnyiNat, Smolyaninov}. In this geometry the surface wave guided along the metallic cone gradually slows down until it is fully localized at the apex\cite{stockman2004, gramotnev2014}. Surprisingly, for the plasmonic vortices (PV), with singularity coinciding with the cone center, the behavior is radically different\cite{pfeiffer1974, schmidt2008, spittel2015}. The sharpening cone leads to the decrease of the effective index and the mode accelerates until it detaches from the surface due to the full momentum matching with the free space. At this specific point an intriguing polarization anomaly can be observed. The radiation emitted from the metal nanotip appears to be fully polarized in one circular state corresponding to the vortex topology. Here we experimentally demonstrate this unique phenomenon and analyze it using recently discovered plasmonic property - the transverse spin\cite{bliokh2015}. The ability to control and analyze the polarization state in nanoscale shall play a pivotal role in nanophotonics, optical encryption and quantum optics. Moreover, local excitation of chiral optical field may be utilized for a single molecule circular dichroism probing\cite{tang2010optical}.

A coupling of the circular polarization handedness to the orbital angular or linear momentum of SPs was previously widely discussed in terms of the ``plasmonic spin-orbit interaction''\cite{gorodetski2008,bliokh2015spin}. 
This interaction resulted in intriguing spin-based phenomena such as the plasmonic spin-Hall effect\cite{shitrit2011optical,gorodetski2012weak}, spin-dependent plasmonic routing\cite{shitrit2013rashba} and guiding\cite{gorodetski2009PRL}, spin-based imaging \cite{capasso2016}, excitation of spin-dependent PVs\cite{gorodetski2008, gorodetski2009PRL} and  spin-dependent far--field beaming\cite{yuri2013, garoli2016}. These phenomena stemmed from a Doppler-like transfer of a longitudinal optical spin (polarization handedness) to the plasmonic orbital angular momentum manifested by its helical phase-front. 
Nevertheless, it was recently shown, that a plasmonic wave could also carry a transverse spin angular momentum (TSAM)\cite{bliokh2012,bliokh2015quantum,bliokh2015, aiello2015} whose role in light-SP coupling might be crucial. The TSAM of the surface wave propagating in $x$ direction on a metal-air interface is given as 
\begin{equation}
  \textbf{s}_{\perp} \propto \frac{Re\textbf{k} \times Im \textbf{k}}{(Re\textbf{k})^2}
\label{TSdef}
\end{equation}
where $\textbf{k} = k_{SP}\mathbf{\hat{x}}+i\kappa \mathbf{\hat{z}}$ is the the complex valued evanescent wave vector, $\kappa = \sqrt{k_{SP}^2 - k_0^2}$, $k_0 = 2\pi/\lambda_0$ is the vacuum wavenumber and $k_{SP}$ is the in-plane plasmonic wavenumber\cite {bliokh2015}.
This transverse spin results from the rotation of the resultant of the vectorial plasmonic field, $\textbf{E}_{SP} = \textit{E}_{p}(\hat{\textbf{z}}-i\chi\hat{\textbf{x}})$ in a transverse plane with respect to the propagation. Remarkably, the TSAM is independent of the polarization and solely arises from the amplitude ratio between the longitudinal and the transverse field components that is directly obtained from Maxwell's equations, $s_{\perp}=\chi =\frac {\kappa}{k_{SP}}$.
Accordingly, $s_{\perp}$ is locked to the SPs propagation direction and can appear with a single handedness. This property has been already utilized for spin-dependent unidirectional plasmonic excitation\cite{miroshnichenko2013, o2014spin}, for nanoparticle tweezing\cite {lee2012role, antognozzi2016, neugebauer2015, canaguier2014} and for study of quantum plasmonic effects\cite{bliokh2015quantum}. 

Although the spin-orbit interaction reported in previous papers referred solely to the longitudinal spin-to-orbital AM transfer we note that considering the transverse AM is essential in non-paraxial systems. Specifically, when an SP mode is guided along a smooth 3D surface and then perfectly impedance-matched to the free space, its TSAM can be fully coupled to a pure circular polarization  (CP) state. Here we experimentally observe a gigantic symmetry breaking in the CP state of light emerging from the tapered nano-cone placed in the center of a plasmonic vortex lens (PVL). One of the circular components experiences almost an order-of-magnitude suppression that results from an adiabatic acceleration of the PV along the nanotip followed by a perfect matching to the far--field. This phenomenon is explained using purely geometric consideration on the TSAM transfer along the tip and shown to be inherent in any smooth 3D SP-guiding system. Therefore the striking importance of the discovered effect in the fundamental physics as well as in a wide field of nano-photonic and quantum applications is evident.   


	\begin{figure}[H]
	\centering
		\includegraphics[width=10cm, keepaspectratio] {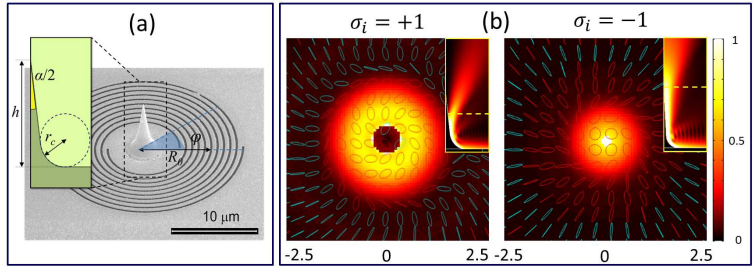}
		\label{fig:Fig1}
		\caption{Anomalous spin emission. (a)- SEM image of the fabricated sample with $m = 2$. The optimized geometrical parameters of the nanotip are $r_c = 1.5\mu m$, $\alpha = 7.5^{\circ}$, $h = 6.3 \mu m$. (b) - Calculated intensity and polarization distribution at the vortex detachment point for different incident spin states. The insets show the intensity flow along the tip and the yellow dashed line represent the height at which the distribution was taken. The scale in (b) is in microns.}
	\end{figure}

The SP launching grating consists of spiral slits engraved in a $300 nm$ metal layer. The spiral radii are given by $R_m (\varphi)= R_0 + m\cdot\varphi/k_{SP}$, where $R_0$ is the smallest radius of the groove, $m$ is the topological order of the spiral and $\varphi$ is the azimuthal angle. The tip is located at the center of the spiral as schematically presented  in \textbf{Figure 1a}. The base curvature was optimized to provide a smooth SP propagation. The specific geometric parameters of the tip are given in the Figure caption. The scanning electron microscope (SEM) image of the fabricated structure is given in \textbf{Figure 1b}. The structure is illuminated from the bottom with CP light whose spin number is denoted as $\sigma_i = +1$ for the right handed and $\sigma_i = -1$ for the left handed state. The incident beam excites a PV whose $E_z$ field component is characterized by a helical phase front\cite {gorodetski2008, gorodetski2009PRL, garoli2016, yuri2013, garoli2016beaming} $exp(il\varphi)$, with the topological charge $l = m+\sigma_{i}$. First, we consider a spiral with $m = 2$ that generates PVs with $l = 3$ or $l = 1$ depending on $\sigma_i$. The propagation of these plasmonic modes along the tip is calculated using COMSOL multiphysics \textregistered  (see the insets of \textbf{Figure 1b}). We note that the $l = 1$ mode beams out exactly at the tip end while the mode with $l = 3$ detaches at some cut-off height. This behavior can be explained by the gradual phase velocity increase of the mode until the full momentum matching to the free space\cite{garoli2016beaming}. We calculate the transverse field distribution slightly above the detachment points (shown by the yellow dashed line in the insets). The complex field values are used to calculate the local polarization ellipse that is graphically presented on the top of the intensity distribution (\textbf{Figure 1b}). The emerging polarization handedness, $\sigma_o = +1$ is shown in red while $\sigma_o = -1$ is in magenta. Note that the emerging modes are both right-handed and very close to the circular state. In other words our system emits $\sigma_o  = 1$ \textit{independently} on the incident handedness. Apparently, most of the previously discussed axially symmetric scattering architectures, such as circular or coaxial apertures were shown to couple PVs to radially polarized beam, that naturally consisted of almost equal amounts of right and left CP\cite{gorodetski2008, gorodetski2009PRL, yuri2013}. Here we link the emission of a single-handed polarization to the TSAM of the plasmonic mode.   


	\begin{figure}[H]
	\centering
		\includegraphics[width=10cm, keepaspectratio] {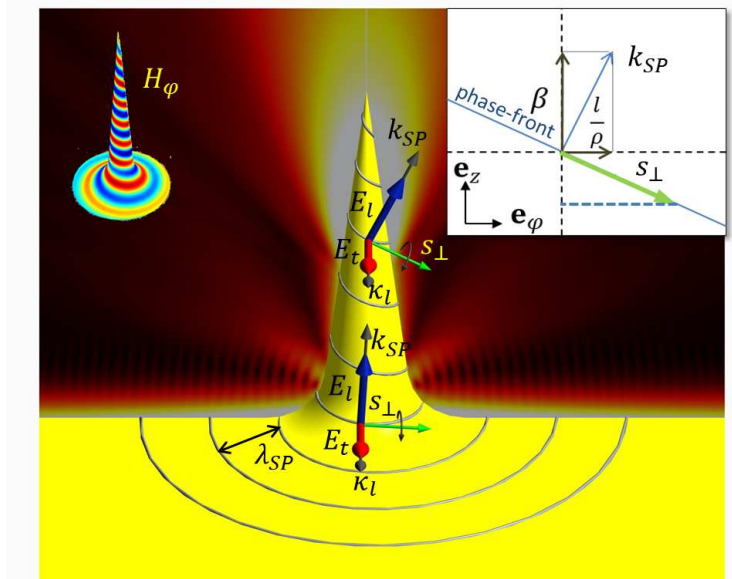}
		\label{fig:Fig2}
		\caption{Geometry used for modeling of the plasmonic mode. The blue line represents the spiralling phase-front of the SP. The $E_l$ and $E_t$ stand for the longitudinal and the transverse components of the plasmonic field, while $k_{SP}$ and $\kappa_l$ denote the real and the imaginary part of the local complex plasmonic wave-vector. The local plasmonic TSAM is denoted as $s_{\perp}$.The  inset on the left is a simulated azimuthal component of the magnetic field $H_{\phi}$. The inset on the right shows the local tangential frame with the mode's propagation vector $\beta$ and the azimuthal wavenumber $l/\rho$.}
	\end{figure}

Therefore we look closer at the geometry of the system which is shown in \textbf{Figure 2}. A spiral phase-front of the PV (represented as the blue line) propagates on a smoothed cone with a local wave vector $k_{SP}$. The transverse and the longitudinal components of the local plasmonic field are denoted in the Figure as $E_t$ and $E_l$, respectively. As can be seen the incidence plane follows the phase-front and gets tilted as the mode propagates upwards. Accordingly, the $z$ projection of the local TSAM (green arrow) grows. To simplify the analysis we treat the conic surface as being comprised of short cylindrical sections of a constant radius $\rho(z)$\cite {spittel2015}.  


	\begin{figure}[H]
	\centering
		\includegraphics[width=10cm, keepaspectratio] {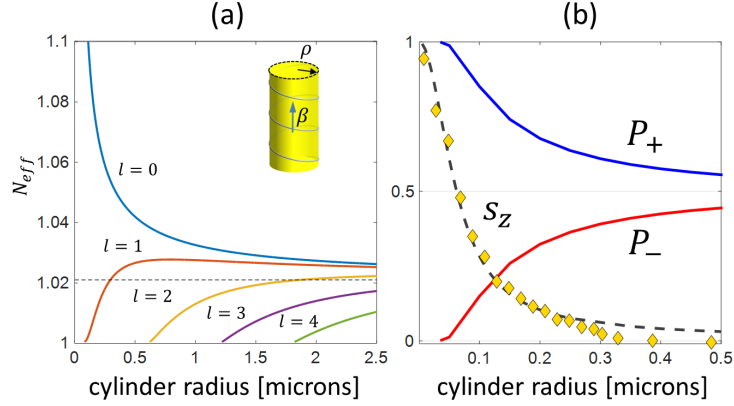}
		\label{fig:Fig3}
		\caption{Calculated dependence of the effective mode index and the emerging spin on the cylinder radius. (a) - Effective index, $N_{eff}$ is calculated using a direct calculation of boundary conditions (solid) and by using a surface impedance matching model (dashed) for different topological charges. (b) - The blue and the red lines show directly simulated helicity of the far--field mode propagated on the surface of the cylinder. The dashed line represents the TSAM of the SP mode propagating on the cylinder calculated using our model. The rhombuses are realistic simulations of the TSAM at different radii of a cone.}
	\end{figure}

In this geometry the plasmonic mode propagation constant $\beta$ can be determined by separately solving the Helmholtz equation in cylindrical coordinates in the dielectric and metal regions, and imposing the continuity of the tangential components of the fields
\cite{pfeiffer1974,stratton2007}. From the calculated mode we then derive the effective refraction index using $N_{eff} = \beta/ k_0$. The dependence of $N_{eff}$ on the cylinder radius is depicted in \textbf{Figure 3a}. The mode with $l = 0$ experiences a darting slowing as the tip radius decreases, which corresponds to the well-known effect of energy localization at the apex\cite {stockman2004, gramotnev2014, de2010nanoscale}. This is the manifestation of the plasmonic ``black hole'' as the energy does not leave the tip, but concentrates around the high index tip apex. For PVs with $l > 0$ the index decreases towards the apex and the modes accelerate up to the free-space phase velocity ($N_{eff} = 1$) where they finally detach.

In a local tangential reference frame $(\mathbf{e}_z, \mathbf{e}_{\varphi})$ the complex plasmonic wave vector is represented as $\mathbf {k}_l = \beta \mathbf{e}_z + \frac{l}{\rho} \mathbf {e}_{\varphi} + i\kappa_l \mathbf {e}_{\rho}$. 
By substituting the $\mathbf{k}_l$ in \textbf{Equation 1} the TSAM of the plasmonic wave can be calculated. In order to study the far--field helicity we consider only the $z$ component of the TSAM,
\begin{equation}
 s_{z} = -\frac{\kappa_l}{k_{SP}^2}\frac{l}{\rho}  
\label{Sz}
\end{equation}
where $\kappa _l = \sqrt{\beta^2 + \left(\frac{l}{\rho}\right)^2-(2\pi/\lambda_0)^2}$ and $k_{SP}^{2} = \beta^2+(l/\rho)^2$. 
The mode detaches from the tip where the effective index becomes unity. The emerging spin at that point is then $s_{z} = -\left[k_0^2(\rho/l)^2+1\right]^{-1}$. 
In \textbf{Figure 3b} we consider the mode $l = 1$ propagating on a cylinder and analyze the transverse fields at some height. We use the calculated fields in the circular basis, $\mathbf{E}_{\pm} = \left|\mathbf{E}\pm i Z_0\mathbf{H}\right|$ ($Z_0$ is the vacuum impedance) to derive the local field helicity current \cite{AielloBerry}, $\mathbf{P}_{\pm} = \pm \epsilon_0 k_0 Im \left[ \mathbf{E}_{\pm}^{\ast} \times \mathbf{E}_{\pm} \right]/\left| \mathbf{E}_{\pm} \right|^2$ and depict its integrated value in \textbf{Figure 3b}. At the same Figure we present the calculated $s_{z}$ (\textbf{Equation \ref{Sz}}) as a function of the cylinder radius by using previously derived $\beta$ values. Finally, we simulate an SP mode propagating on the metallic cone with realistic parameters and directly calculate the TSAM in the vicinity of the cone at different heights. Its integrated values are shown in \textbf{Figure 3b} as yellow rhombuses. We note that the values of $s_{z}$ calculated at the real tip fully correspond to the ones extracted from the modal analysis of the cylinder. Moreover, it is clearly visible that for the small radius both, the far-field ellipticity $P_{+}$ and the $s_{z}$ become unity, which indicates the emission of a pure circular polarization. On the other hand, for large radii the $s_{z}$ tends to zero. This indicates that if  the field was emitted from this point its polarization would contain equal amounts of right handed and left handed CP. The latter behavior is expected from the metal due to its non-duality and was widely discussed in\cite{fernandez2012}. Obviously, the described effect of anomalous spin selectivity can be expected from all helical plasmonic modes. This can be understood from \textbf{Equation \ref{Sz}} with the real cutoff radius values, where $s_z$ rapidly goes to unity as a function of $l$.          


	\begin{figure}[H]
	\centering
		\includegraphics[width=10cm, keepaspectratio] {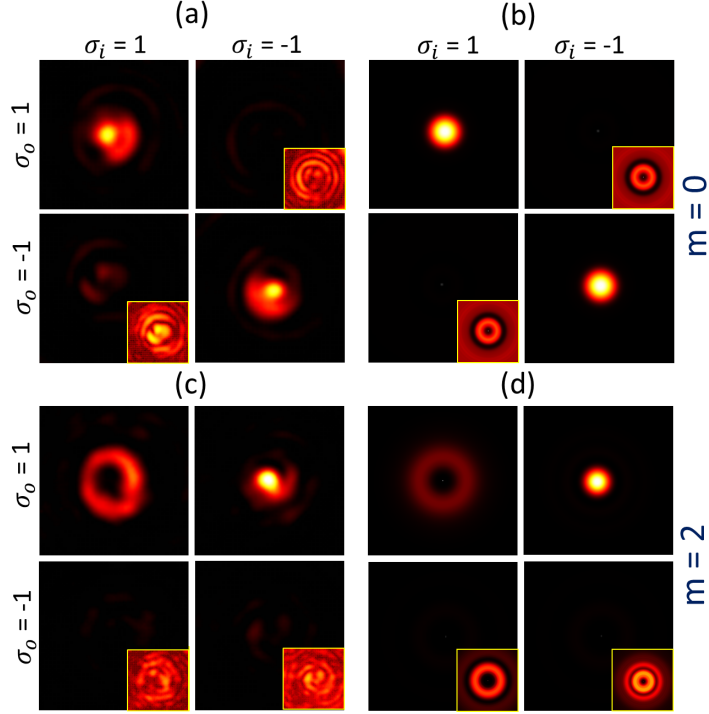}
		\label{fig:Fig4}
		\caption{Anomalous spin far--field emission. (a) and (c) - Experimentally measured far--field distributions for different polarization state combinations with (a) - $m = 0$ and (c) - $m = 2$. (b) and (d) - Corresponding simulated distributions. Insets show distributions in logarithmic scales.}
	\end{figure}

To visualize the effect of a single-handed helicity emission we use two different ways to generate a plasmonic vortex with $l = 1$. For this we have fabricated nano-structures with two types of the grating: $m = 2$ and $m = 0$ with the same conic tip in the center. The samples were illuminated from the back side by CW pigtail laser at $\lambda_0 = 785nm$. The incident and the emerging spin states were controlled using the setup described in the supplementary section. The distributions captured by the camera are presented in \textbf{Figure 4a} and \textbf{Figure 4c} for each  polarization combination ($\sigma_i, \sigma_o$). In \textbf{Figure 4b} and \textbf{Figure 4d} we show the calculated intensities.   

In the case of $m = 0$, a plasmonic helical mode with $l = \pm 1$ is excited as expected from the AM conservation in a circular symmetry\cite{gorodetski2008,gorodetski2009PRL}. The PV then propagates up to the end of the cone where it emits a purely circularly polarized light with $\sigma_0 = \sigma_i$ as can be elucidated from the strong contrast between the diagonal and anti-diagonal distributions in \textbf{Figure 4a}. Nevertheless, by using a spiral with $m = 2$ a PVs with $l = 1,3$ are generated depending on the incident spin. According to our model the emerging $s_z$ should approach unity and a \textsl{single CP} is expected in the far--field.  The experimental and the calculated intensity distributions in this system are shown in \textbf{Figure 4b}. Here the upper row intensity clearly exceeds the ones in the lower panels. Note, that both presented structures generated PV with $l = 1$ at the right CP state although they were initially excited by different circular components. The result of the two experiments clearly shows that the emerging light helicity is locked to the PV handedness, $\sigma_o = sgn(l)$, due to the non-trivial 3D geometry of the guiding surface. The overall average ratio of the maximum distribution intensities of the distinct states was found to be 7 as opposed to the ratio of 20 expected from the simulations. We associate this discrepancy to a tiny step at the basis of the fabricated nano-tip, resulting in a weak scattering of light.   

In summary, we experimentally presented and theoretically analyzed the helicity-locking during a beaming of a plasmonic mode from an adiabatically tapered metallic cone. This behavior was attributed to the coupling of the plasmonic TSAM to the far--field due to the smooth geometry of the tip and the effective mode acceleration. This phenomenon, however, obeys the angular momentum conservation prescribed by the 3D geometry of our system. We note that in contrast with previous works \cite{canaguier2014,neugebauer2015,antognozzi2016,lee2012role,o2014spin,lefier2015,miroshnichenko2013} our system is the first to show a coupling of TSAM to the far--field polarization while the propagation direction before and after the beaming is not changed. This might be important for various applications in nanophotonics and optics communications. In addition we beleive that this plasmonic effect has a fundamental character and can be generalized to other systems and scales. One of the examples in which our demonstrated effect may shed some light is the spontaneous helicity locking of a radiation emitted by a distant star collapsing into a black hole\cite{wiersema2014}.  

\begin{acknowledgement}
This work was supported by the Ministry of Science Technology $\&$ Space, Israel.
\end{acknowledgement}

\begin{suppinfo}

\subsection*{Fabrication} 

The fabrication of the samples is based on a procedure described by De Angelis et al\footnotemark. The principle relies on FIB-generated secondary-electron lithography in optical resists and allows the preparation of high aspect ratio structure with any 3D profile. The final structure comprises of a $6.2 \mu m$ high base-smoothed gold tip on a 150 nm gold layer where PVLs are milled. In order to prepare such a complex architecture a multi-step fabrication process have been optimized. First of all a 5 / 23 nm Ti / Au bilayer has been deposited, by means of sputtering, on a 100 nm thick Si3N4 membrane. On this conductive layer, s1813 optical resist has been spun at 1500 rpm and soft-baked at $90^{\circ}$C for 8 minutes. The resist thickness of $11 \mu m$ is achieved by tuning the concentration, spinning time and velocity. On the back of the membrane a thin layer of silver (about $10 nm$) is then deposited by means of sputtering in order to ensure the necessary conductibility of the sample for the successive lithographic step. The membranes are then patterned from the backside using a Focused Ion Beam (Helios Nanolab600, FEI company), operated at $30 keV$ (current aperture: $80pA$, dwell time: $500 \mu s$). The tip-like shape has been obtained by patterning successive disks with decreasing diameter and correcting the dose applied for every disk, thus resembling the expected tip profile. (To note that the first milled disk present a high thickness (around $80 nm$) that will be filled in the successive metallic growth). Due to the high dose of low-energy secondary electrons induced by the interaction between the ion beam and the sample, a $30 nm$ thick layer of resist, surrounding the milled disks, becomes highly cross-linked and insoluble to most solvents. After patterning, the sample is developed in acetone, rinsed in isopropanol and dried under gentle $N_2$ flow. The back side silver layer has been then removed by means of rapid $HNO_3$ rinse. At this stage we get a high dielectric tip surrounded by a metallic substrate. Since we need a base-smoothed tip on a $150 nm$ thick gold layer, an additional layer of metal has been grown of the substrate by means of galvanic deposition ($0.12 $Amp DC). The galvanic layer is grown up to the tip base so ensuring a very smooth geometry. After the galvanic deposition, a $40 nm$ thick layer of gold is deposited by sputtering the sample, tilted  $60^{\circ}$ with respect to the vertical and rotated, guaranteeing an isotropic coating on both the sidewalls and the base. (In order to avoid any possible direct transmittance from the tips, the back of them has been filled, by means of electron beam induced deposition, with a $200 nm$ thick layer of platinum). Finally, in order to prepare the sample with the desired m-order PVL surrounding the tip, a FIB milling process has been performed on the sample creating the spiral gratings without affecting the quality of the metallic tip. To maximize the structure symmetry we use a set of $m$ spirals, each one is rotated by $2\pi/m$ with respect to the next one, in such a way that the radial distance between the two adjacent grooves stays $\lambda_{SP} = 2\pi/k_{SP}$, thus improving the coupling of normally impinging light.
 
\subsection*{Optical Setup} 
The transmission far--field measurements are performed using free-space optical setup consisting of alluminating CW pigtail laser operating at $\lambda_0 = 785nm$. The laser beam is collimated and polarized by a set of vertically oriented LP followed by a QWP rotated at $45^{\circ}$. The light is pre-focused on the back side of the PVL by means of 20X microscope objective (NA = 0.25). The imaging was performed by using an infinity-corrected 60X objective (NA = 0.85) followed by a 100 mm tube lens and an additional 1.5X magnification telescope. The emerging spin-state was tuned by an additional QWP-LP set placed in the image path.  The resulting image was captured by a PIXELINK CMOS industrial camera (PL-B771U, MONO 27, 1280X1024).    

\footnotetext{ De Angelis,F. et al. Nano Lett. \textbf{13}, 3553--8 (2013).}

\end{suppinfo}


\end{document}